\documentclass[final,1p,times]{elsarticle} 

\usepackage{graphicx}
\usepackage{amssymb} 
\usepackage{amsthm} 
\usepackage{lineno}
\usepackage{mediabb}

\journal{Nuclear Physics A} 

\begin{document}

\begin{frontmatter} 

\title{Detailed HBT measurements with respect to the event plane and collision energy in Au+Au collisions at PHENIX}

\author{Takafumi Niida for the PHENIX\fnref{col1} Collaboration}
\fntext[col1] {A list of members of the PHENIX Collaboration and acknowledgements can be found at the end of this issue.}
\address{University of Tsukuba, 1-1-1 Tennoudai, Tsukuba, Ibaraki 305-8571, Japan}


\begin{abstract} 
The azimuthal dependence of 3D HBT radii relative to the event plane gives us 
information about the source shape at freeze-out. It also provides  
information on the system's evolution by comparing it to the initial 
source shape.
In recent studies, higher harmonic event planes and flow have been measured at RHIC and the LHC, 
which result primarily from spatial fluctuations of the initial density across the collision area. 
If the shape caused by initial fluctuations still exists at freeze-out,
the HBT measurement relative to higher order event plane may show these features.

We present recent results of azimuthal HBT measurements relative to 
$2^{nd}$- and $3^{rd}$-order event planes in Au+Au 200 GeV collisions with the PHENIX experiment. 
Recent HBT measurements at lower energies will be also shown and compared with the 200 GeV result. 
\end{abstract} 

\end{frontmatter} 


\section{Introduction}
HBT measurements provide information on the space-time 
evolution of the particle emitting source in relativistic heavy ion collisions.
The azimuthal dependence of 3D HBT radii with respect to an event plane gives us 
information on the source shape at freeze-out. It also provides information 
on the system's evolution by comparing it to the initial source shape.
The higher harmonic flow ($v_{3}$, $v_{4}$, etc) of particles
has recently been measured at RHIC and the LHC. It is primarily due to
the spatial fluctuation of the initial density of the collision area. 
A hydrodynamic model calculation \cite{voloshin} reports that the shape of the initial fluctuations 
resulting in a triangular component of the initial shape may be preserved until freeze-out. 
HBT measurements relative to a higher-order event plane may reveal this feature. 


\section{Azimuthal HBT with respect to the $2^{nd}$-order event plane}
\begin{figure}[htbp]
\begin{center}
\includegraphics[width=0.55\textwidth,angle=90]{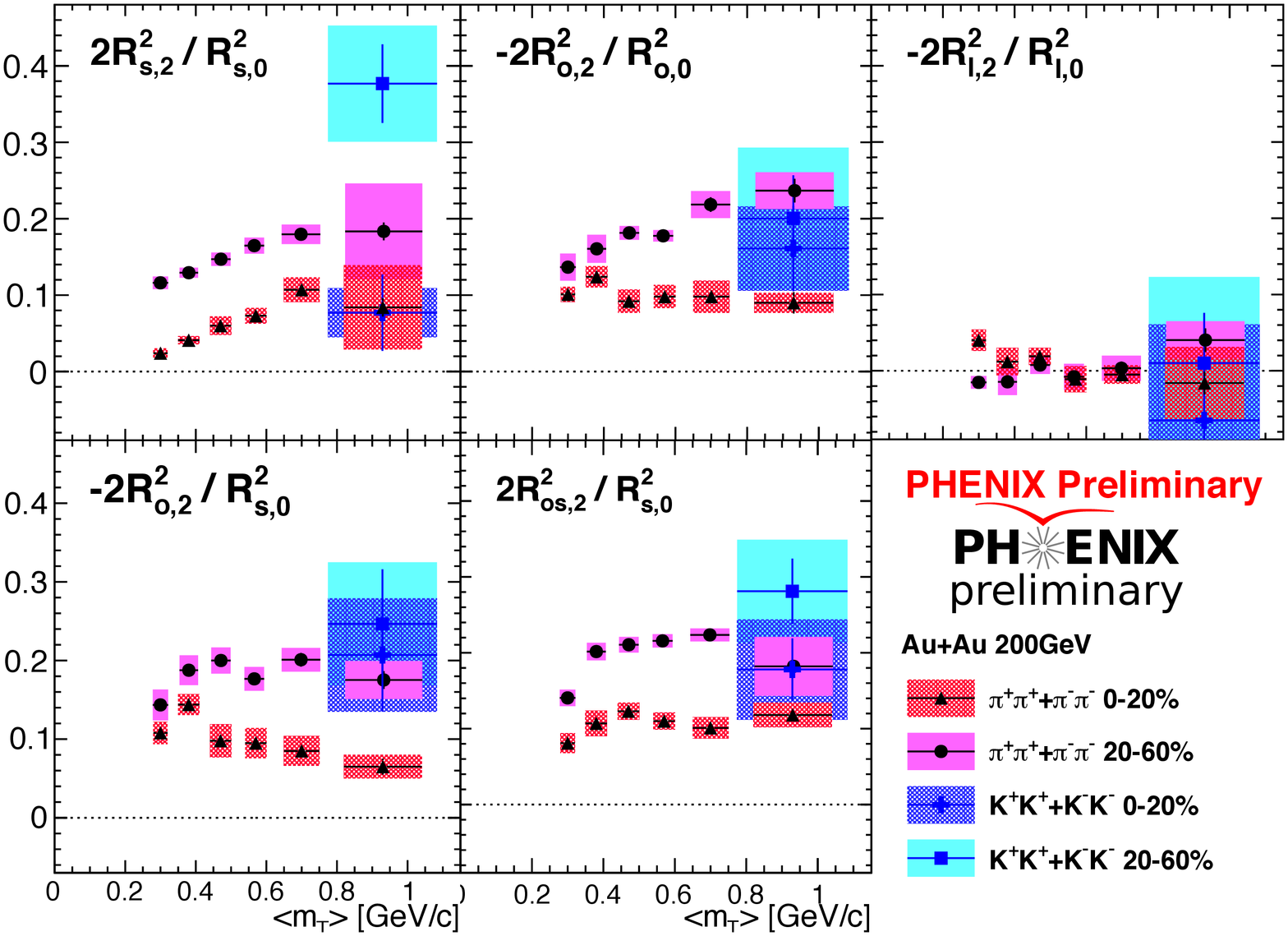}
\end{center}
\caption{Relative amplitude of azimuthal HBT radii for charged pions and kaons 
with respect to $2^{nd}$-order event plane in Au+Au 200 GeV collisions.}
\label{fig:hbt_e2}
\end{figure}

Azimuthal HBT radii with respect to the $2^{nd}$-order event plane have been measured 
for charged pions and kaons in $\sqrt{s_{\rm NN}}$ = 200 GeV Au+Au collisions at PHENIX \cite{hbt}. 
It was found that the final eccentricity of kaons, which is defined as
$\varepsilon_{2,final}$ = 2$R_{s,2}^{2}$/$R_{s,0}^{2}$ \cite{bw}, is larger than that of pions and almost the same 
as the initial eccentricity. However, since HBT radii show a transverse mass ($m_{T}$) dependence and 
the average $m_{T}$ of pions and kaons are different, the $m_{T}$ dependence needs to be considered 
in the comparison of the final eccentricities.
Figure \ref{fig:hbt_e2} shows the relative amplitude of the azimuthal HBT radii for charged pions and kaons as a function of 
$\langle m_{T} \rangle$ for two centrality bins. The left top panel corresponds to the final eccentricity; there is
still a difference between pions and kaons in non-central collisions even at the same $\langle m_{T} \rangle$. 
This difference may be due to different cross-sections of pions and kaons.
The relative amplitudes of $R_{o}$ and $R_{os}$ at low $m_{T}$ in the most central collisions have 
finite values although the final eccentricity is almost close to zero. 
This result may indicate a temporal variation of the emission duration of particles because 
$R_{o}$ and $R_{os}$ contain temporal information in addition to geometrical information.

\section{Azimuthal HBT with respect to $3^{rd}$-order event plane}
Fluctuations in the initial geometry of the heavy ion collisions, which is considered to be 
the origin of higher harmonic flow, may be preserved until freeze-out. 
Triangular flow, $v_{3}$, is known to have a weak centrality dependence, 
while the initial triangularity calculated within a Glauber model has a pronounced centrality dependence \cite{esumi}. 
Triangularity at freeze-out is determined by the initial triangularity, $v_{3}$, the expansion time, and so on,
and will provide detailed information on the space-time evolution.

\begin{figure}[t]
\begin{center}
\includegraphics[width=0.34\textwidth,angle=90]{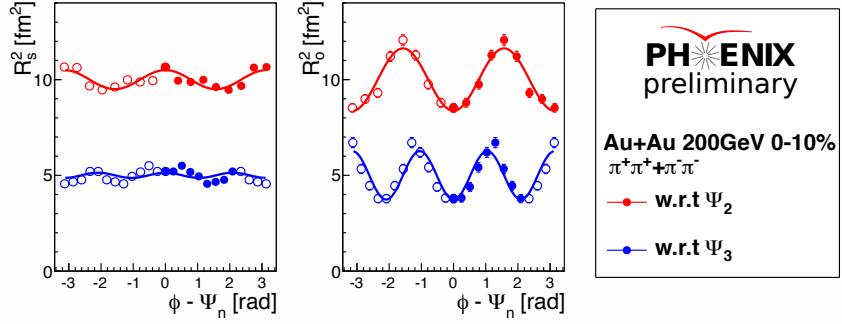}
\end{center}
\caption{The azimuthal dependence of $R_{s}$ and $R_{o}$ for charged pions 
with respect to $2^{nd}$- and $3^{rd}$-order event plane in Au+Au 200 GeV collisions,
where the averages of radii with respect to $\Psi_{2}$ and $\Psi_{3}$ are set to 10 and 5 fm$^{2}$ respectively. }
\label{fig:psi3}
\end{figure}

Figure \ref{fig:psi3} shows $R_{s}$ and $R_{o}$ for charged pions 
as a function of the azimuthal angle with respect to $2^{nd}$- and $3^{rd}$-order event planes
($\Psi_{2}$ and $\Psi_{3}$) 
in 0-10\% in Au+Au 200 GeV collisions, where the averages of the radii with respect to $\Psi_{2}$ and $\Psi_{3}$
are set to 10 and 5 fm$^{2}$ respectively. Filled symbols show measured data points, and 
open symbols are the same data points reflected around $\phi-\Psi_{n}=0$. The solid lines depict the fit functions 
$R_{\mu, 0}$+2$R_{\mu, n}$cos$(n(\phi-\Psi_{n}))$ ($\mu$ = $s,$ $o$).
The values of $R_{s}$ show a very weak oscillation with respect to both $\Psi_{2}$ and $\Psi_{3}$, 
while $R_{o}$ clearly exhibits stronger oscillation.
The oscillation strength of $R_{o}$ relative to $\Psi_{3}$ is comparable to that relative to $\Psi_{2}$.
The oscillation of $R_{o}$ may carry information about the duration of the emission. 
The $\Psi_{2}$ dependence may indicate that the emission duration is different in-plane versus out-of-plane 
since $R_{s}$ shows a weak oscillation and the source shape is thought to be close to a circle. 
The $\Psi_{3}$ dependence may be also due to the difference of emission duration between 
different azimuthal directions. 
However, the flatness of $R_{s}$ doesn't necessarily mean that the source shape is not
triangular but circular \cite{voloshin}. 
It may be difficult to imagine that the emission duration has such strong variations in azimuth 
relative to different event planes.
The oscillation of $R_{o}$ may reflect not only the emission duration but also the depth of 
the elliptical or triangular source. 

\begin{figure}[h]
\begin{center}
\includegraphics[width=0.41\textwidth,angle=90]{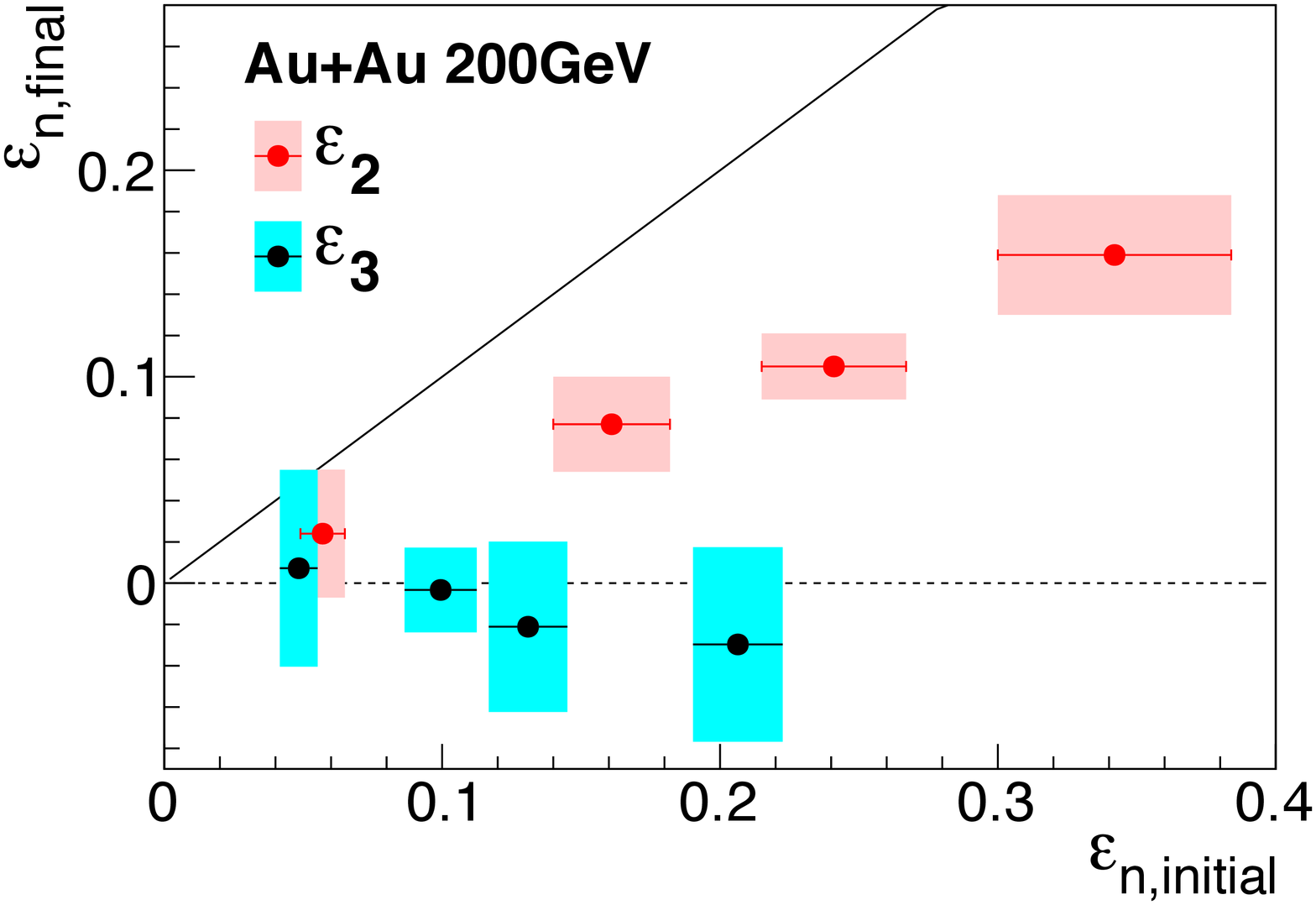}
\end{center}
\caption{Final $\varepsilon_{n}$ as a function of initial $\varepsilon_{n}$ in Au+Au 200 GeV collisions. 
Initial $\varepsilon_{n}$ is calculated in a Glauber model.}
\label{fig:e2e3}
\end{figure}

As mentioned before, the final eccentricity is defined as $\varepsilon_{2,final}$ = 2$R_{s,2}^{2}$/$R_{s,0}^{2}$.
Here, we define $\varepsilon_{3,final}$ as 2$R_{s,3}^{2}$/$R_{s,0}^{2}$.
Figure \ref{fig:e2e3} shows $\varepsilon_{n,final}$ as a function of $\varepsilon_{n,initial}$,
where $\varepsilon_{n,initial}$ is calculated using a Glauber model.
Note that $\varepsilon_{3,final}$ doesn't represent the final triangularity
because there is no higher harmonic anisotropy in the Gaussian approximation for a static source.
However, it will mean any triangular component of homogeneity region in a expanding source.
The observed $\varepsilon_{3,final}$ doesn't seem to exhibit any centrality dependence and is zero within systematic uncertainties.

\section{Low energies at PHENIX}
\begin{figure}[t]
\begin{center}
\includegraphics[width=0.34\textwidth, clip, angle=90, trim=50 20 30 20]{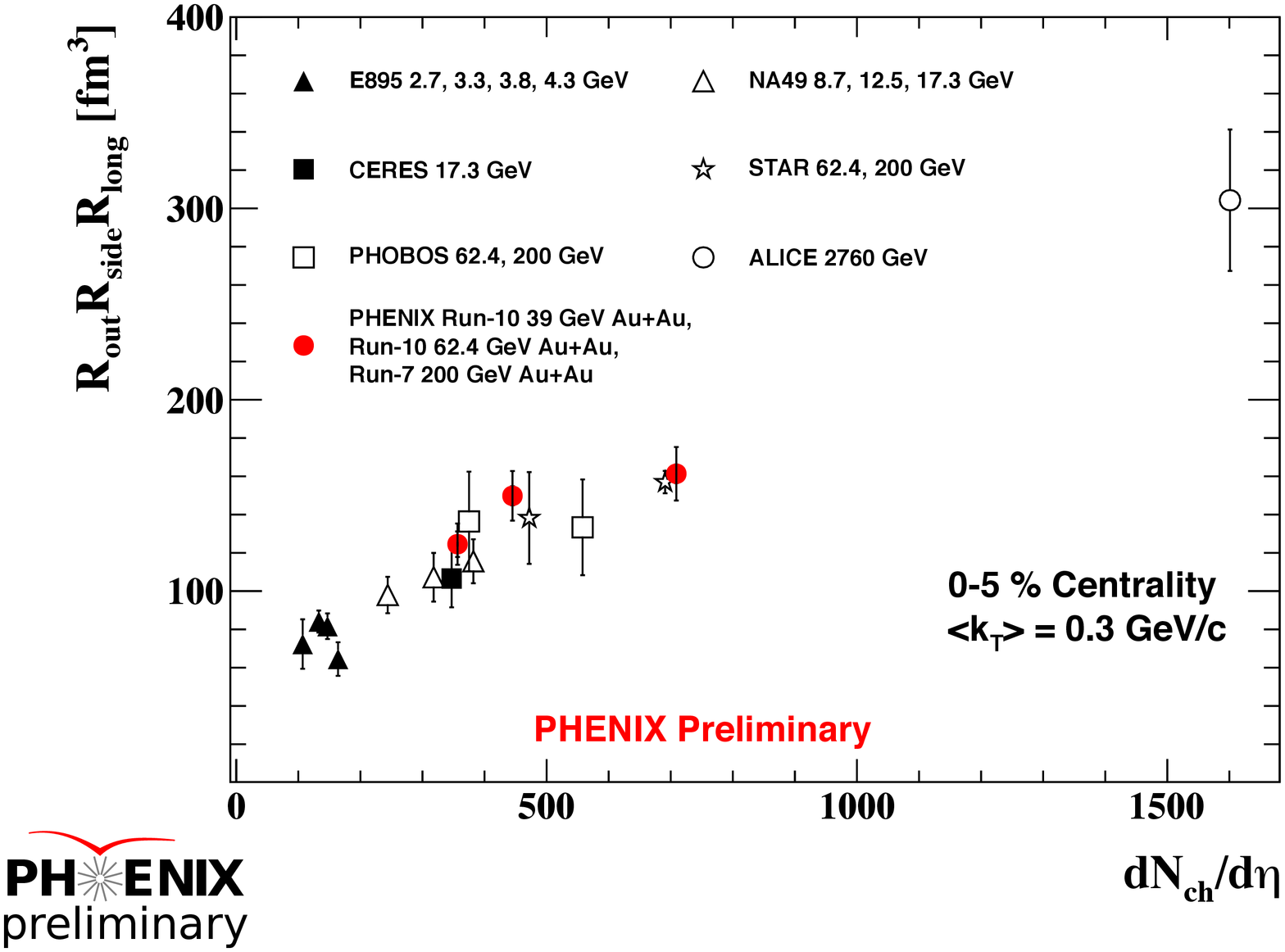}
\end{center}
\caption{The product of pion 3D HBT radii as a function of charged particle multiplicity density.}
\label{fig:volume}
\end{figure}

The RHIC beam energy scan program has been conducted 
in order to explore the critical point between the QGP and hadron gas phases in the QCD phase diagram,
The centrality and $m_{T}$ dependence of HBT radii were measured for 39, 62 and 200 GeV collision energies.
We observe no significant change beyond systematic errors across the three energies.
Figure \ref{fig:volume} shows the product of 3D HBT radii for charged pions as a function of 
charged multiplicity density. PHENIX results are compared with results at different energies and collision species at AGS\cite{e895}, SPS\cite{na49,ceres}, RHIC\cite{phobos,star} and LHC\cite{alice}. The product of 3D HBT radii, which represents the volume of 
the homogeneity region, from PHENIX is consistent with the global trend.

\section{Summary}
The latest results of azimuthal HBT measurements with respect to the $2^{nd}$-  and $3^{rd}$-order event planes 
are presented. The difference of final eccentricity for charged pions and kaons is seen even in the same 
$m_{T}$ region, which may indicate a shorter freeze-out time of kaons due to a lower cross section. 
The azimuthal dependence of pion HBT radii relative to $\Psi_{3}$ has been measured and the oscillation of
$R_{o}$ is clearly seen. This may be indicative of the temporal variation of the emission duration or the depth of the source
with triangular shape. These results will provide new constraints on theoretical models and new insight into
the space-time evolution of the system.


\end{document}